\documentclass[twocolumn,prl]{revtex4}
\usepackage{graphicx}
\usepackage{amsmath}

\begin{document}
\title{A Nearly Scale Invariant Spectrum of Gravitational Radiation from Global Phase Transitions}
\author{Katherine Jones-Smith$^1$}
\author{Lawrence M. Krauss$^{1,2}$}
\author{Harsh Mathur$^1$}
\affiliation{$^1$CERCA, Department of Physics, Case Western Reserve University,10900 Euclid Avenue, Cleveland OH 44106-7079 \\ $^2$ Also Department of Astronomy, Case Western Reserve University}

\begin{abstract}

Using a large N sigma model approximation we explicitly calculate the power spectrum of gravitational waves arising from a global phase transition in the early universe and we confirm that it is scale invariant, implying an observation of such a spectrum may not be a unique feature of inflation.  Moreover, the predicted amplitude can be over 3 orders of magnitude larger than the naive dimensional estimate, implying that even a transition that occurs after inflation may dominate in Cosmic Microwave Background polarization or other gravity wave signals.

\end{abstract}

\maketitle
In 1992  \cite{lmk1992} it was proposed on dimensional grounds that a scalar field which undergoes a symmetry breaking phase transition can give rise to a scale-invariant spectrum of gravitational waves by virtue of causality: the correlation length of the field is of order the horizon size, as the horizon grows, uncorrelated regions come into contact, and the field releases energy as it relaxes.  
  
Following the detection of both temperature and polarization fluctuations in the Cosmic Microwave Background (CMB), there is now great interest in the possibility of using CMB measurements to probe for a nearly scale invariant gravitational wave spectrum that is widely considered to be the ``smoking gun'' of inflation \cite{starobkrausswmap}.  Long wavelength primordial gravitational radiation would affect the polarization of the CMB \cite{pol}.  

Given the importance therefore of a possible observation of a gravitational wave signature in the CMB, and proposed missions capable of observing CMB polarization such as CMBPol and Planck
it is imperative
to go beyond dimensional arguments and explore the detailed nature of the spectrum of radiation
produced by another mechanism such as that in \cite{lmk1992} from which the inflationary spectrum would
need to be discriminated.   This is the purpose of the present work.

Our analysis proceeds in two stages. First we analyze the ordering of 
a symmetry breaking scalar field evolving in a flat FRW background. Then
we compute the gravitational radiation produced by the relaxation of the 
scalar field. The estimates given in \cite{lmk1992} assumed a two component scalar 
field in a Mexican hat potential. However the two component case presents
special complications due to the possibility that strings will form and frustrate
the ordering. For this reason we work with an $N$-component field and work 
in the sigma model limit. In other words we impose the constraint 
\begin{equation}
\sum_{\alpha = 1}^N  \phi_{\alpha}^2 = \eta^2
\label{eq:constraint}
\end{equation}
on $N$ otherwise free massless fields. Here $\eta$ is the vacuum expectation value
of the field. Due to the constraint the equations of
motion for this system are non-linear and in general can only be solved
numerically. However it has long been known in the condensed matter
literature on ordering kinetics that the problem becomes tractable in the
large $N$ limit; that large $N$ method was adapted to the study of 
global phase transitions in cosmology 
by Turok and Spergel \cite{turokspergel} whose analysis
we will follow. 

A key approximation in the large $N$ solution is to replace the trace of the stress-tensor
with its spatial average in the sigma model equation of motion. Thus we set about 
solving 
\begin{equation}
\frac{\partial^2\phi_{\alpha}}{\partial \tau^2}+ \frac{2}{a}\frac{\partial a}{\partial \tau}\frac{\partial \phi_{\alpha}}{\partial \tau} - \nabla^2 \phi_{\alpha} = -\frac{1}{\eta^2}\mathcal{T}(\tau)\phi_{\alpha}
\label{eq:spatialaverage}
\end{equation}
where $\tau$ is conformal time, $a$ is the scale factor, and $\mathcal{T}(\tau)$
is the spatial average of trace of the sigma field 
stress tensor, $a^2 T^{\mu}\hspace{0.01mm}_{\mu} 
({\mathbf r}, \tau)$ \footnote{For the sigma model in an FRW background the trace
of the stress tensor $a^2 T^{\mu}\hspace{0.01mm}_{\mu} = 
\sum_{\alpha = 1}^{N} [ ( \nabla \phi_{\alpha} )^2 - 
( \partial \phi_{\alpha}/\partial \tau)^2 ]$.} .
For simplicity we first consider the case that the scale factor varies as $a \propto 
\tau^{\beta}$. The solution is constructed self-consistently. 
We posit that $\mathcal{T} = - T_0 \eta^2/\tau^{2}$ and find
 \begin{equation}
 \phi_{\alpha}({\mathbf k}, \tau) = A_{\alpha}\tau^{(1/2)-\beta}J_{\nu}(k\tau)
 \label{eq:bessel}
 \end{equation}
where $\phi_{\alpha}({\mathbf k}, \tau)$ is the spatial Fourier transform of the
sigma model field, $A_\alpha$ is to be determined by the initial condition and
the order of the Bessel function is given by $\nu^2 = T_0 + (\beta - 1/2)^2$;
both $T_0$ and $\nu$ will be determined self-consistently below.

To model the initial symmetric state we take the sigma field to be completely
disordered at time $\tau_0$. 
To regulate the initial field distribution we introduce
a real space cubic lattice with lattice constant $d$, and choose ``white
noise'' initial conditions where the field components at each lattice site are 
taken to be independent Gaussian random variables. The variance of the 
distribution is chosen so that the constraint eq (\ref{eq:constraint}) 
at each site is satisfied on average;
the deviations from the mean vanish in the large $N$ limit. 
Since the Fourier amplitudes of the sigma field 
evolve independently according to 
eq (\ref{eq:bessel}) they remain independent Gaussian random variables.
Their correlators at later times can be evaluated in terms of their 
correlators at the initial time $\tau_0$.

Finally we impose self-consistency by requiring the average stress
tensor trace ${\mathcal T}(\tau)$ computed from our solution $\phi_{\alpha}({\mathbf k},
\tau)$ to equal the postulated form $- T_0 \eta^2/ \tau^2$. This 
requirement \footnote{As a practical matter it is more convenient to impose the
self-consistency condition that the solution satisfy the spatially averaged constraint 
eq (\ref{eq:constraint}) than
to compute the trace of the stress tensor. Remarkably the two self-consistency
conditions can be shown
to be equivalent \cite{turokspergel}. At this stage we also take the continuum limit
$d \rightarrow 0$ and $\tau_0 \rightarrow 0$. Self-consistency requires that we set
$(d/\tau_0)^3 = 2 \pi^2 \Gamma( \beta + 1/2 ) \Gamma( 2 \beta + 3/2 )/2^{2 \beta} 
\Gamma( \beta ) [ \Gamma( \beta + 2 ) ]^2$.}
determines that $\nu = 1 + \beta$ and $T_0 = 3 \beta + \frac{3}{4}$.

In summary, in the large $N$ limit, the two-point correlators
of the sigma model field are given by
$\langle \phi_{\alpha} ({\mathbf k}, \tau_1) \phi_{\gamma} ({\mathbf p}, \tau_2) \rangle
= (2 \pi)^3 \delta_{\alpha \gamma} \delta ({\mathbf k} + {\mathbf p})
C( k, \tau_1, \tau_2 )$ and
\begin{eqnarray}
C( k, \tau_1, \tau_2 ) & = &
8 \pi^2 \frac{ \eta^2 }{N} 
\frac{ \Gamma( \beta + 1/2 ) \Gamma( 2 \beta + 3/2 )}{\Gamma( \beta )}
\nonumber \\
& & \times \frac{ \tau_1^{1/2 - \beta} \tau_2^{1/2 - \beta} }{k^{2 + 2 \beta} }
J_{1 + \beta} ( k \tau_1 )
J_{1 + \beta} ( k \tau_2 ).
\nonumber \\
\label{eq:correlator}
\end{eqnarray}
Four-point and higher correlators may be computed by use of Wick's 
theorem. 

Experience with ordering dynamics suggests 
for example that the equal-time two-point correlator should have
the scaling form $ C(k, \tau, \tau) = \tau^{\alpha} f( k \tau^z )$.
Indeed eq (\ref{eq:correlator}) has this form with dynamical exponent 
$z = 1$ and $ \alpha = 3 $. That the weight of a mode, $| \phi_{\alpha} ({\mathbf k},
\tau) |^2$, grows as $ \tau^3$ before it comes into the horizon ($\tau \ll 1/k$) can
be understood as a consequence of the constraint eq (\ref{eq:constraint}). 
Initially all modes have the same weight, but as the universe expands, the
field orders on the horizon scale, 
and short wavelength modes that have already come into the horizon 
decay. To satisfy the
constraint, spectral weight shifts into the long wavelength modes that lie
inside a sphere of radius $1/\tau$ in wave-vector space. Since the total 
weight is conserved, the weight of these modes grows as $\tau^3$. 

So far we have assumed that the scale factor varies as a power law
$a \propto \tau^{\beta}$. This is valid deep in the radiation and matter
dominated eras with $\beta = 1$ and 2 respectively. (Note that these
are the relevant eras for CMB polarization studies.)  To model the crossover
between the two eras we take the scale factor to be
\begin{equation}
a \propto \frac{\tau}{\tau^{\ast}}+(\frac{\tau}{2\tau^{\ast}})^2
\label{eq:crossover}
\end{equation}
where $\tau^{\ast}$ is the crossover time. 
We define a time dependent exponent $b(\tau) \equiv (\tau/a) \partial a/\partial \tau$.
For a pure power law scale factor this coincides with the exponent $\beta$, and for the 
crossover form,
$b(\tau)$ has the asymptotic values 1 for $ \tau \ll \tau^{\ast}$
and 2 for $\tau \gg \tau^{\ast}$. Eq (\ref{eq:spatialaverage}) is no longer
soluble in closed form when $a$ is given by eq (\ref{eq:crossover}), but 
since $b(\tau)$ varies slowly, 
it is an excellent approximation to work with the solution for fixed
$\beta$ derived above 
and simply make judicious replacements of $b(\tau)$ in place of $\beta$.
Turok and Spergel \cite{turokspergel} have verified the accuracy of this ``adiabatic''
approximation numerically. Note that in this case the sigma field correlators
depart weakly from a rigorous scaling form since $b$ varies slowly with $\tau$.

We now turn to the calculation of the gravitational radiation. We decompose the
metric into $g_{\mu \nu} = \eta_{\mu \nu}(\tau) + h_{\mu \nu}$ where $\eta_{\mu \nu}
= a^2 (\tau) ( d \tau^2 - d x^2 - d y^2 - d z^2 )$ 
is the flat space FRW metric and $h_{\mu \nu}$ is the perturbation
due to the evolving scalar field. As customary in linearized relativity we
decompose the perturbation into scalar, vector and traceless tensor parts via
$h_{00} = 2 a^2 \Phi$, $h_{0i} = - a^2 w_i$ and $h_{ij} = - 2 a^2 s_{ij} + 
2 a^2 \psi \delta_{ij}$. It is most convenient to use 
the transverse gauge $ \partial_i w_i = 0$ and $\partial_i s_{ij} = 0$
and to work with the spatial Fourier transform of the perturbation.
In this representation, for each wave-vector there are two transverse
degrees of freedom whose amplitudes are conventionally written 
$h_+({\mathbf p}, \tau)$ and $h_{\times}({\mathbf p}, \tau)$. 
These amplitudes obey the linearized Einstein equations
\begin{equation}
\frac{\partial^2}{\partial \tau^2} h_+ +
\frac{2}{a} \frac{\partial a}{\partial \tau} \frac{\partial h_+}{\partial \tau}
+ p^2 h_+ = T_+
\label{eq:lineinstein}
\end{equation}
and a similar equation for $h_{\times}$. For the wavevector ${\mathbf p}$
along the $z$-axis the non-vanishing components of the metric perturbation
are related to the amplitudes via
$s_{11} ({\mathbf p}, \tau) = - s_{22} ({\mathbf p}, \tau) = h_+$ 
and $s_{12} ({\mathbf p}, \tau) = s_{21} ({\mathbf p}, \tau) = h_{\times}$. For 
other orientations of the wave-vector, $s_{ij}$ can be computed by
rotating the components of the tensor appropriately. The gravitational
modes $h_+$ and $h_{\times}$ are driven by the stress in the sigma model
field. For ${\mathbf p}$ along the $z$-axis the driving terms are related
to the components of the sigma-field stress tensor via
$T_+({\mathbf p}, \tau) = 4 \pi G [T_{11} ({\mathbf p}, \tau) - 
T_{22} ({\mathbf p}, \tau)]$ and $T_{\times} ({\mathbf p}, \tau) = 8 \pi G
T_{12} ({\mathbf p}, \tau)$\footnote{The relevant components of the
sigma field stress tensor $T_{11} = (1/2)\sum_{\alpha=1}^{N} 
(\partial \phi_{\alpha}/\partial \tau)^2 +
(\partial \phi_{\alpha}/\partial x)^2 -
(\partial \phi_{\alpha}/\partial y)^2 -
(\partial \phi_{\alpha}/\partial z)^2$, and 
$T_{12} = \sum_{\alpha = 1}^{N} 
(\partial \phi_{\alpha}/\partial x)(\partial \phi_{\alpha}/\partial y)$.}. 
For other orientations of the wave-vector
the source terms can be constructed by appropriately rotating the 
components of the sigma-field stress-tensor.

Before we embark on a full solution it is useful to gain some 
qualitative understanding of eq (\ref{eq:lineinstein}). In Minkowski
space $a = 1$ and gravitational mode amplitudes respond as simple
harmonic oscillators. In the FRW case there is an additional time-dependent
damping produced by the expansion of the Universe. For $a \propto \tau^{\beta}$
and in the absence of external driving ($T_+ = 0$) it is easy to show that
the oscillations of the gravitational wave amplitude decay as 
$1/a$ for late times $\tau \gg 1/p$; in other words, modes redshift and
decay once they come into the horizon. For the case that a mode is driven by
a power law source, $T_+ \propto \tau^n$, it is easy to show that for short 
times, $\tau \ll 1/p$, a previously undisturbed gravitational mode grows
as $h_+ \propto \tau^{n+2}$. This is because for short times the mode 
amplitude behaves essentially like a driven free particle. For the case 
of interest, where the source is the sigma field stress, we expect that
$ T_+({\mathbf p}, \tau) \approx 0$ in the long time limit $\tau \gg 1/p$ since
the sigma field is ordered on sub-horizon scales. In other words a gravitational
mode should evolve freely, without external driving, once it comes into the horizon.
For short times, $\tau \ll 1/p$, the driving term 
$ T_+({\mathbf p}, \tau) \sim 1/\sqrt{\tau}$; this follows simply
from the scaling form of the sigma field correlators and power-counting. 
Thus we infer that $| h_+({\mathbf p}, \tau)|^2 \propto \tau^3$ for short
times, $\tau \ll 1/p$. For long times, $| h_+({\mathbf p}, \tau)|^2 \propto 1/a^2$
Finally for $a \propto \tau^{\beta}$ it is easy to see
that, since the sigma field correlators have a scaling form, so too will
$|h_+({\mathbf p}, \tau)|^2$. This translates into the observation that
the gravitational strain spectrum (defined below) is not a separate function
of $p$ and $\tau$ but of the product $p \tau$. Thus the spectrum will be scale
invariant in the sense the term is commonly used in the literature. 

The linearized Einstein eq (\ref{eq:lineinstein}) is soluble by conventional
Green's function methods \cite{morse}:
\begin{equation}
h_+({\mathbf p}, \tau) = \frac{1}{a(\tau)}
\int_{0}^{\tau} d \tau_1 G_{{\mathbf p}}( \tau, \tau_1 )
a(\tau_1)
T_+({\mathbf p}, \tau_1).
\label{eq:green}
\end{equation}
For $a \propto \tau^{\beta}$ the Green's function is given by
\begin{equation}
G_{{\mathbf p}}(\tau, \tau_1) =  
p \tau \tau_1 
[ j_{{\beta - 1}} (p \tau_1) n_{{\beta - 1}} (p \tau) -
n_{{\beta - 1}} (p \tau_1) j_{{\beta - 1}} (p \tau) ];
\label{eq:moregreen}
\end{equation}
here $j_{\beta-1}$ is the spherical Bessel function and $n_{\beta -1}$
the spherical Neuman. For $a$ given by eq (\ref{eq:crossover}) we evaluate
the Green's functions using the adiabatic approximation, replacing $\beta$
with $b(\tau)$ and $b(\tau_1)$ appropriately. The gravitational strain
spectrum is defined as
\begin{equation}
\frac{2 \pi^2}{p^3} P(p, \tau ) \delta ({\mathbf k} + {\mathbf p}) = 
\langle h_+({\mathbf p}, \tau)  h_+({\mathbf k}, \tau) + 
h_{\times} ({\mathbf p}, \tau) h_{\times}({\mathbf k}, \tau) \rangle.
\label{eq:power}
\end{equation}
Eqs (\ref{eq:correlator}), (\ref{eq:green}) and (\ref{eq:moregreen}) provide
all the ingredients required to compute the gravitational strain spectrum. 
Performing this calculation, making use of Wick's theorem to evaluate
the required four-point sigma field correlator, we arrive at our final 
result 
\begin{eqnarray}
P(p, \tau) & = & 2048 \pi^5 \frac{G^2 \eta^4}{N} \frac{1}{a(\tau)^2} p^3
\int_{0}^{\infty} d k \int_{0}^{\pi} d \theta \sin^5 \theta   
\nonumber \\
&  \times &  k^4 (k^2 + p^2 - 2 k p \cos \theta)^{-1}
F(k, p, \theta, \tau)^2;
\nonumber \\
F(k, p, \theta, \tau) & = & \int_{0}^{\tau} d \tau_1 
\tau_1^{1 - 2 b} k^{- b} (k^2 + p^2 - 2 k p \cos \theta)^{-\frac{b}{2}} 
\nonumber \\
& \times & 
\frac{ \Gamma( b + \frac{1}{2}) \Gamma( 2 b + 
\frac{3}{2} ) }{\Gamma( b )}
G_{{\mathbf p}}(\tau, \tau_1) a (\tau_1)
\nonumber \\
& \times &
J_{1 + b}( k \tau_1 ) J_{1 + b} ( 
\sqrt{k^2 + p^2 - 2 k p \cos \theta} \tau_1 ).
\nonumber \\
\label{eq:monster}
\end{eqnarray}
This result simplifies if $ a \propto \tau^{\beta}$: 
then $b \rightarrow \beta$ and we may set $\tau = 1$ since the power
becomes a function of the product $p \tau$ (as can be seen by suitable
rescaling). Fig 1 shows the strain spectrum evaluated in this way
for the case of matter domination $\beta = 2$. 
For small $p \tau$ we have derived the asymptotic result
\begin{equation}
P(p, \tau) = C_{\beta} \frac{G^2 \eta^4}{N} (p \tau)^3
\label{eq:small}
\end{equation}
consistent with the behavior deduced on physical grounds above.
The constant $C_{\beta}$, which gives the approximate power at $p \tau=1$, is $ \approx 2500 $ for $\beta = 1$ and 1000 for $\beta = 2$. 
For large $p \tau$ the numerically calculated fall off is
consistent with the expected $1/a^2$ form but we have not yet 
confirmed this by asymptotic analysis. Right at the peak (which occurs at
$ p\tau \approx 3.7 $) we find the power goes as
$\kappa G^2 \eta^4/N$, in agreement with the estimate of ref \cite{lmk1992},
but factors of $2 \pi$ remarkably conspire to make the numerical factor $\kappa =11,600 $. Thus the
gravitational radiation produced by this mechanism is over four 
orders of magnitude larger than was estimated there.
Modes that enter the 
horizon in the matter dominated era satisfy $p \tau^{\ast} \ll 1$; for 
these modes Fig 1 is a very good 
approximation for any $\tau \gg \tau^{\ast}$, {\em i.e.}, the entire relevant
history of the Universe.

Our key findings are summarized in Figs 1 and 2. Fig 1 shows the 
power spectrum of gravitational radiation $P(p, \tau)$ calculated in 
an approximation that is valid in the matter dominated era.
Here $p$ is the magnitude of the gravitational wavevector 
and $\tau$ is conformal time. As we have shown, the matter dominated era the
power is a function of the product $p \tau$.  Fig 1 shows 
that for modes of fixed $p$ the power peaks at $ \tau \sim 1/p$ 
when the modes come into the horizon. For later times the modes
redshift and consequently the power falls off as $1/\tau^4$. For
earlier times the power grows as $\tau^3$. 
Fig 1 may also be interpreted as a plot of power {\em vs.}
wave vector for fixed time. Then, the power
is seen to peak at wavevectors on the horizon scale $p = 1/\tau$
and to drop
off larger and smaller scales. 

Fig 2 shows a plot of the
exact power, eq (\ref{eq:monster}), as a function of conformal time $\tau$
at wave-vector $p = 1/\tau$, over a much more complete range of time.
Fig 2 shows that the spectrum is indeed scale
invariant over the range of times $\tau \gg \tau^{\ast}$ and that it
approaches the value expected for pure matter domination, with a jump at matter-radiation crossover time, and a plateau for $\tau \ll \tau^{\ast}$ at the value for pure
radiation domination.

\begin{figure}
\scalebox{0.6}{\includegraphics{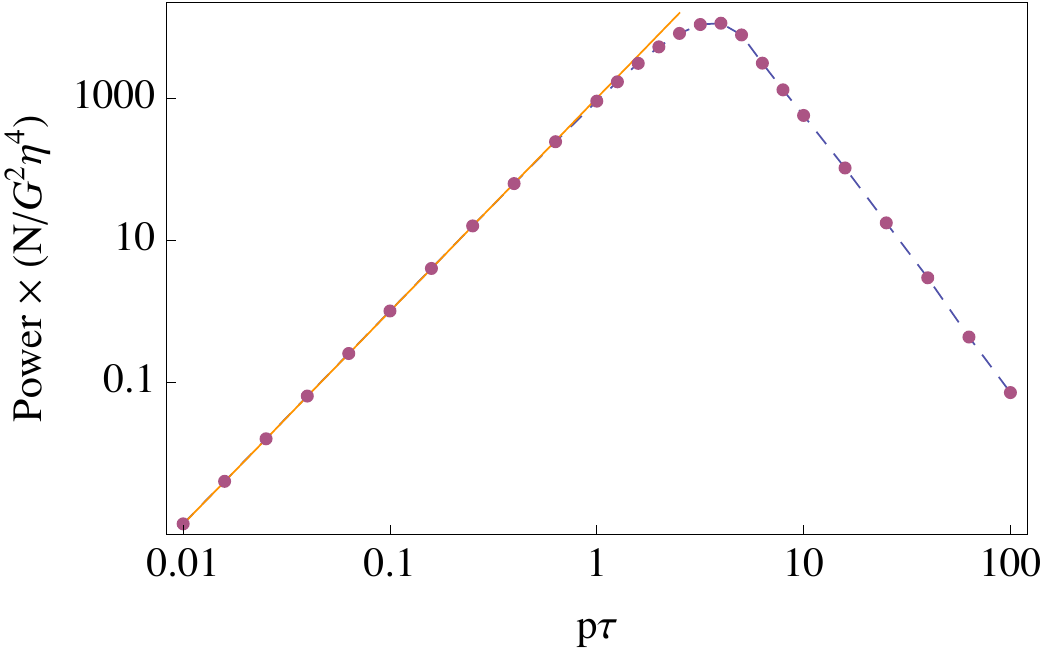}}
\caption{Plot of gravitational radiation power
vs $p \tau$ where $p$ is the wave-vector and $\tau$ is
the conformal time, computed assuming the
scale factor appropriate for matter domination. 
The plot may be interpreted to show the variation in power
with time for modes of a fixed wave-vector or the variation in
power with wave-vector for fixed time. The points are obtained by
numerical integration of eq (\ref{eq:monster}). The
dashed curve is a guide to the eye. The solid line shows
an asymptotic approximation $C_{\beta} (p \tau)^3$ valid for
small $p \tau$. The numerical data are consistent with a $1/(p \tau)^4$
tail for large $p \tau$. The peak value 11,600 occurs at
$p \tau \approx 3.7$.}
\label{fig:crossover}
\end{figure}

\begin{figure}
\scalebox{0.6}{\includegraphics{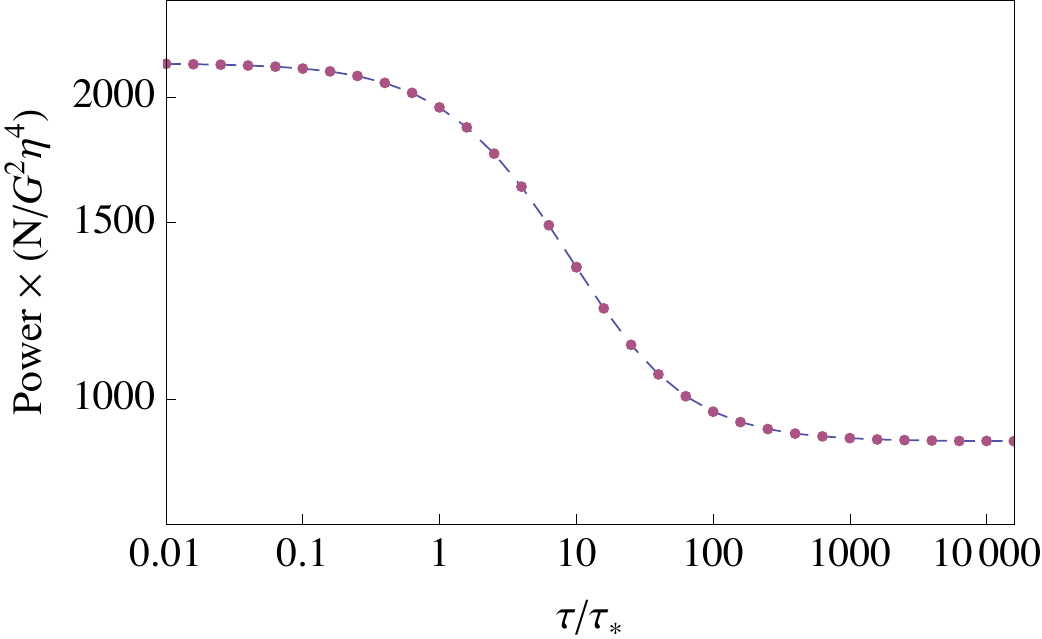}}
\caption{Plot of the gravitational power vs $\tau/\tau_*$ at wavevector
$p = 1/\tau$. $\tau_*$ is the time of crossover from radiation
to matter domination. The spectrum is flat for $\tau \gg \tau_*$
and again for $\tau \ll \tau_*$; thus except at the crossover,
it is essentially scale invariant. The points are calculated by
numerical integration of eq (\ref{eq:monster}); the dashed curve
is a guide to the eye.}
\label{fig:crossover}
\end{figure}

These results not only confirm expectations based on dimensional grounds, they provide
significant motivation to further explore in detail the possible observational consequences of 
the gravitational wave spectrum predicted here.  The surprisingly large magnitude of the power in gravitational waves produced by this mechanism does not require inflation not to have occurred for the spectrum generated by such a global transition to be observable.  Even a phase transition below the scale of inflation might produce a signature that could compete with or overwhelm that generated during inflation.  As a result some observable difference in the resulting polarization signal will be necessary if we hope to claim such a signal may provide unambiguous evidence that inflation occurred in the early universe. Work is now in progress to calculate the CMB polarization signature due to this radiation in order to seek out possible distinctions in the signatures for the different alternatives. 

Another figure of merit useful for experimental purposes is 
$\Omega_{{\rm gw}} (p, \tau_{\rm now})$, the present energy density
per log frequency interval of subhorizon 
modes normalized to the critical density.
It is given by $\Omega_{{\rm gw}} (p, \tau_{\rm now}) = p^2
P(p, \tau_{\rm now})/ 6 H^2$ where $H$ is the present Hubble constant.
At frequencies that
can be detected by advanced LIGO, LISA and millisecond pulsars, both inflation
and this mechanism predict a flat spectrum for
$\Omega_{{\rm gw}} (p, \tau_{\rm now})$, with relative strengths given by:
$$ R=  {3  \kappa \over 4 } \left[{ \eta \over {N^{1/4}v}}\right]^4 \approx 10^4 \left[\eta \over {N^{1/4} v} \right]^4 $$
 where $v^4 = V_0$ is 
the energy density during inflation.  Once again, depending upon relative energy scales, this signature could dominate over that due to inflation. 
 
As a final note, while we expect that the general features of the resulting spectrum are not dependent on the large $N$ approximation which allowed an exact analytical form to be determined, it may also be  desirable to study the
case $N = 2$ and $N=3$ cases. Here the formation of
topological defects (strings and monopoles) will also influence the dynamics
of the symmetry breaking field, leaving an imprint on the spectrum of 
gravitational radiation.

We acknowledge discussions with Irit Maor and Tanmay Vachaspati.



\end{document}